\documentclass[final,a4paper]{amsart}
\usepackage{amsmath,amssymb,amsthm}
\usepackage[margin]{fixme}
\usepackage{setspace}
\usepackage{subfigure}
\usepackage[final]{graphicx}
\usepackage{enumerate}         

\ifx\@draft\@undefined \else \usepackage{showkeys}\fi

\newtheorem{thm}{Theorem}[section]
\newtheorem{cor}[thm]{Corollary}
\newtheorem{lem}[thm]{Lemma}

\theoremstyle{definition}
\newtheorem{defn}[thm]{Definition}
\newtheorem{cond}[thm]{Condition}

\newtheorem{app}[thm]{Approximation}
\theoremstyle{remark}
\newtheorem{rem}[thm]{Remark}
\newtheorem{example}[thm]{Example}

\def\JELname{\textbf{Journal of Economic Literature Classification}\enspace}
\def\JEL#1{\par\addvspace\medskipamount{\rightskip=0pt plus1cm
\def\and{\ifhmode\unskip\nobreak\fi\ $\cdot$
}\noindent\JELname\ignorespaces#1\par}}

\numberwithin{equation}{section}

\newcommand{\norm}[1]{\left\Vert#1\right\Vert}

\newcommand{\set}[1]{\left\{#1\right\}}
\newcommand{\Ind}[1]{\mathbf{1}_{\left\{#1\right\}}}
\newcommand{\RR}{\mathbb{R}}

\newcommand{\PP}{\mathbb{P}}
\newcommand{\CC}{\mathbb{C}}
\newcommand{\NN}{\mathbb{N}}

\newcommand{\cF}{\mathcal{F}}

\newcommand{\cL}{\mathcal{L}}

\renewcommand{\Re}{\mathrm{Re}\,}

\newcommand{\E}[1]{\mathbb{E}\left[#1\right]}                     
\newcommand{\Econd}[2]{\mathbb{E}\left[\left.#1\right|#2\right]}        
\newcommand{\Varcond}[2]{\mathbb{V}\mathrm{ar}\left[\left.#1\right|#2\right]}        

\begin{document}
\title[Asymptotic and Exact Pricing of Options on Variance]{Asymptotic and Exact Pricing of Options on Variance}
\author{Martin Keller-Ressel}
\address{ETH Z\"urich, Departement Mathematik, R\"amistrasse 101, CH-8092 Z\"urich, \linebreak \indent Switzerland}
\email{kemartin@math.ethz.ch}
\author{Johannes Muhle-Karbe}
\address{ETH Z\"urich, Departement Mathematik, R\"amistrasse 101, CH-8092 Z\"urich, \linebreak \indent Switzerland}
\email{johannes.muhle-karbe@math.ethz.ch}
\date{November 19, 2010.}

\thanks{Financial support by the National Centre of Competence in Research `Financial Valuation and Risk Management' (NCCR FINRISK), Project D1 (Mathematical Methods in Financial Risk Management) is gratefully acknowledged. The NCCR FINRISK is a research instrument of the Swiss National Science Foundation.\\[5pt]
We thank Richard Vierthauer for valuable discussions on the regularity of Laplace transforms and Marcel Nutz for comments on an earlier version. We are also very grateful to two anonymous referees, whose helpful comments significantly improved the present article.}

\begin{abstract}
We consider the pricing of derivatives written on the \emph{discretely sampled} realized variance of an underlying security. In the literature, the realized variance is usually approximated by its continuous-time limit, the quadratic variation of the underlying log-price. Here, we characterize the small-time limits of options on both objects. We find that the difference between them  strongly depends on whether or not the stock price process has jumps. Subsequently, we propose two new methods to evaluate the price of options on the discretely sampled realized variance. One of the methods is approximative; it is based on correcting prices of options on quadratic variation by our asymptotic results. The other method is exact; it uses a novel randomization approach and applies Fourier-Laplace techniques. We compare the methods and illustrate our results by some numerical examples.
\end{abstract}

\maketitle

\section{Introduction}
Consider a discounted asset $S=S_0\exp(X)$ and a time-interval $[0,T]$ subdivided into $n$ intervals of equal length with boundary points $t_j = j \frac{T}{n}$ for $j=1,\ldots, n$. The corresponding (annualized) \emph{realized variance} of $X$ over $[0,T]$ is then defined as
\begin{equation}\label{Eq:realized_variance}
RV^X_n(T) = \frac{1}{T}\sum_{j=1}^{n} \log(S_{t_j}/S_{t_{j-1}})^2 = \frac{1}{T}\sum_{j = 1}^n{\left(X_{t_j} - X_{t_{j-1}}\right)^2}.
\end{equation}
There exists a considerable number of financial instruments that are based on realized variance as an underlying (see, e.g., \cite{Buhler2006,carr.lee.08} or \cite[Chapter~11]{Gatheral2006} for an overview). Well-known examples are variance and volatility swaps, as well as \emph{puts} and \emph{calls on realized variance} with payoffs $(K-RV_T)^+$ resp.\ $(RV_T-K)^+$. By market convention, the length $t_j - t_{j-1}$ of a single interval typically corresponds to one business day for these derivatives (see \cite{Buhler2006,lee2010}). For puts and calls, the notion \emph{At-The-Money} (henceforth \emph{ATM}) refers to choosing the strike $K$ equal the to swap rate, which in turn equals the expectation $\E{RV^X_n(T)}$ under the pricing measure.

Given a stochastic model for $S$ resp.\ $X$, the standard approach to pricing options on realized variance is to approximate realized variance by
\begin{equation}
 \label{Eq:approximation}
RV_n^X(T) \approx \frac{1}{T}[ X, X]_T,
\end{equation}
where $[X,X]$ is the quadratic variation of the log-price $X$. This approximation is motivated by the fact that -- for fixed $T$ -- realized variance \eqref{Eq:realized_variance} converges to $\frac{1}{T}[X,X]_T$ in probability as the number of subdivisions $n$ tends to infinity (cf., e.g., \cite[Theorem I.4.47]{js.03}). The advantage of this approach is that for many stochastic processes, the quadratic variation is a well-studied object. For example, as recently shown by \cite{Kallsen2009}, the characteristic function of the quadratic variation in any affine stochastic volatility model\footnote{The class of affine stochastic volatility models includes exponential L\'evy models, the Heston model with and without jumps, and many stochastic time-change models.} can be computed as the solution of a generalized Riccati differential equation, such that in many cases methods based on Fourier-Laplace inversion (cf. \cite{Carr1999,raible.00}) can be applied to compute option prices efficiently. Moreover, using quadratic variation in place of realized variance, allows -- at least in diffusion models -- for elegant replication arguments, such as the the representation of a variance swap as an infinite portfolio of European options (see \cite{neuberger.1992}).

The quality of the approximation \eqref{Eq:approximation}, or more precisely, the speed of convergence of realized variance to quadratic variation as the number of subdivisions $n$ increases has been studied extensively in the econometric literature. Barndorff-Nielsen and Shephard \cite{barndorff.shephard.01}, for example, obtain a central limit law for the difference between realized variance and quadratic variation, scaled by the square root of $n$, which holds in a large class of stochastic volatility models (compare also \cite{jacod.08} for more general results).

However, an important difference from the econometric setting to the setting of variance options is that the sampling frequency $n$ for such options can not be chosen freely, but is determined by market convention. As mentioned above, daily sampling is the typical case. From an econometric point of view, this sampling frequency would most likely be considered insufficiently low to produce an acceptable estimate of quadratic variation over most reasonable time horizons $T$. For variance options, though, not pointwise estimation under the physical measure is the goal, but the accurate risk-neutral pricing and hedging of options with specific payoffs. Therefore it is not clear to what extent findings from econometrics can be transferred. For this very reason several articles have considered the quality of the approximation \eqref{Eq:approximation} purely from the point of view of option pricing. B\"uhler \cite{Buhler2006} and Sepp \cite{sepp.08}, resp.\ Broadie and Jain \cite{broadie.jain.08} find via Monte-Carlo simulation resp.\ analytically that the approximation \eqref{Eq:approximation} for daily sampled realized variance works very well for claims with linear payoffs, like variance swaps. On the other hand, B\"uhler \cite{Buhler2006} observes that ``\emph{while the approximation of realized variance via quadratic variation works very well for variance swaps, it is not sufficient for non-linear payoffs with short maturities. The effect is common to all variance curve models (or stochastic volatility models, for that matter).}'' In particular, he presents some numerical examples based on call options on realized variance in the Heston model, that indicate that the approximation by quadratic variation notably diverges from the true value for short maturities (cf.\ \cite[p.\ 128]{Buhler2006}). This leads to the following questions considered in the present study:

\begin{enumerate}
\item  To what extent is it indeed true that quadratic variation is not a good proxy for realized variance, when pricing short-dated options with non-linear payoffs?
\item How can options on the realized variance be valuated more accurately?
\end{enumerate}

The remainder of the article is organized as follows.  We first consider exponential L\'evy models in Section~\ref{Sec:levy} and compare the small-time limits of options on quadratic variation and on realized variance. Based on the results we propose a new method to approximatively evaluate prices of options on realized variance. In Section~\ref{Sec:semimartingale} we show that the results on exponential L\'evy models can be transferred without modification to general semimartingale models. In Section~\ref{Sec:exact} we propose a second -- exact -- pricing method for options on realized variance, which is based on Fourier-Laplace methods. We present numerical examples and compare the two methods in Section~\ref{Sec:Example} and then conclude with some suggestions for future research.

 \section{Small-time asymptotics in exponential L\'evy models}\label{Sec:levy}

 In this section, we derive the small-time asymptotics for options on variance in exponential L\'evy models. That is, we suppose the asset price process is modeled as $S = S_0 \exp(X)$ for a L\'evy process $X$. The latter is assumed to be square-integrable, such that the variance swap rate is always finite. The L\'evy process $X$ will be characterized through its L\'evy-Khintchine triplet $(b,\sigma^2,F(dx))$ with respect to the truncation function $h(x) = x$ or, equivalently, by its \emph{L\'evy exponent}, i.e., the function
$$\psi(u)= ub+\frac{1}{2}u^2 \sigma^2 + \int (e^{ux}-1-ux)F(dx), \quad u \in i\mathbb{R},$$
for which $\E{e^{uX_t}}=\exp(t\psi(u))$. We can decompose $X$ as
 \[X_t = bt + \sigma W_t + L_t,\]
 where $W$ is a standard Brownian motion and $L$ is an independent centered pure-jump L\'evy process.

 \subsection{Options on quadratic variation}

 First, we consider the simpler case of options written on (annualized) quadratic variation.

\begin{thm}\label{thm:qvlimit}
Let $X$ be a square-integrable L\'evy process with L\'evy-Khintchine triplet $(b,\sigma^2,F(dx))$ and suppose the payoff functions $g_T: \mathbb{R}_+ \to \mathbb{R}$, $T \geq 0$ are continuous, uniformly bounded, and satisfy $||g_T-g_0||_{\infty}\to 0$ as $T \to 0$. Then
\[\lim_{T\to 0}\E{g_T(\tfrac{1}{T}[X,X]_T)}=g_0(\sigma^2).\]
\end{thm}

The proof is based on the following auxiliary result, which is shown in the appendix.

\begin{lem}\label{lem:auxqv}
Denote by $L$ the pure-jump component of $X$. Then we have
$$\tfrac{1}{T}[L,L]_{T} \to 0\ \mbox{a.s.} \quad  \mbox{as } T \to 0.$$
\end{lem}

\begin{proof}[Proof of Theorem \ref{thm:qvlimit}]
Evidently,
\begin{align*}
 &\left|\E{g_T(\tfrac{1}{T} [X,X]_T)}-g_0(\sigma^2)\right| \\
 &\qquad \leq \E{|g_T(\tfrac{1}{T}[X,X]_T)-g_0(\tfrac{1}{T}[X,X]_T)|}+\E{|g_0(\tfrac{1}{T}[X,X]_T)-g_0(\sigma^2)|}.
 \end{align*}
By dominated convergence, the first term converges to zero, because we have $||g_T-g_0||_{\infty}\to 0$ as $T \to 0$. Likewise, dominated convergence and the continuity of $g_0$ imply that the second term also converges to zero as $T \to 0$, because $\frac{1}{T}[X,X]_T=\sigma^2+\frac{1}{T}[L,L]_T\to \sigma^2$ a.s., by Lemma \ref{lem:auxqv}.
\end{proof}

\subsection{Options on Realized Variance}
The analogue of Theorem \ref{thm:qvlimit} for options on the discrete realized variance \eqref{Eq:realized_variance} reads as follows:

\begin{thm}\label{thm:rvlimit}
Let $X$ be a square-integrable L\'evy process with L\'evy-Khintchine triplet $(b,\sigma^2,F(dx))$ and suppose that the payoff functions $g_{n,T}: \mathbb{R}_+ \to \mathbb{R}$, $T \geq 0, n \in \NN$ are uniformly bounded and satisfy $||g_{n,T}-g_{n,0}||_{\infty} \to 0$ as $T \to 0$ for each $n$. Then
\begin{equation}\label{Eq:limit}
 \lim_{T \to 0} \E{g_{n,T}(RV^X_n(T)} = \E{g_{n,0}(Y_n)},
\end{equation}
where $Y_n$ has gamma distribution with shape parameter $n/2$ and scale parameter $2 \sigma^2/n$.
\end{thm}

Note that the distribution of the limiting random variable $Y_n$ is determined solely by the number $n$ of sampling dates and the (annualized) variance $\sigma^2$ of the continuous part of $X$. The jump part of the process can only affect the value indirectly through the payoff functions $g_{n,T}$, which usually depend on the swap rate and hence on the specification of $X$. For the most important case of calls and puts on variance, it will turn out in Section \ref{sec:discussion} below that the final result indeed depends also on the jump part, namely through the (annualized) variance $v^2=\int x^2 F(dx)$ of the jump measure. For ATM calls and puts, the small-time limit will be determined completely by the value of $v^2$.

To prove Theorem \ref{thm:rvlimit}, we need the following two lemmas. The proof of the first one is similar to the proof of Lemma~\ref{lem:auxqv} and can be found in the appendix.
\begin{lem}\label{lem:auxrv}
Denote by $L$ the pure-jump component of $X$. Then we have
\[RV_n^L(T) \to 0\ \mbox{a.s.} \quad  \mbox{as } T \to 0.\]
\end{lem}

\begin{lem}\label{Lem:conditional_chisq}
Let $\cL = \sigma(L_t; 0\le t \le T)$ be the $\sigma$-algebra generated by the pure-jump component $L$ of $X$.
Then conditionally on $\cL$, the rescaled realized variance $\frac{n}{\sigma^2} RV^X_n(T)$ follows a non-central chi-square distribution, with $n$ degrees of freedom and noncentrality parameter
\[\lambda(T) = \left(\frac{b}{\sigma}\right)^2T + \frac{2b}{\sigma^2} L_T + \frac{n}{\sigma^2}RV_n^L(T).\]
\end{lem}

\begin{proof}
Conditionally on $\cL$, the independent random variables $X_{t_j}-X_{t_{j-1}}$, $j=1,\ldots,n$ are normally distributed with
\begin{align*}
 \Econd{X_{t_j}-X_{t_{j-1}} }{\cL} &= bT/n + L_{t_j}-L_{t_{j-1}},\\
 \Varcond{X_{t_j}-X_{t_{j-1}}}{\cL} &= \sigma^2 T/n.
\end{align*}
The rescaled realized variance $\frac{n}{\sigma^2} RV^X_n(T)$ is therefore non-central chi-square distributed with $n$ degrees of freedom, conditionally on $\cL$. The noncentrality parameter of the distribution is given by
\begin{align*}
\lambda(T)&=\sum_{j=1}^n{\frac{\Econd{X_{t_j}-X_{t_{j-1}}}{\cL}^2}{\Varcond{X_{t_j}-X_{t_{j-1}}}{\cL}}}\\
&= \left(\frac{b}{\sigma}\right)^2T + \frac{2b}{\sigma^2} L_T + \frac{n}{\sigma^2 T}\sum_{j=1}^n (L_{t_j}-L_{t_{j-1}})^2,
\end{align*}
as claimed.
\end{proof}

\begin{proof}[Proof of Theorem \ref{thm:rvlimit}]
Denote by
\begin{equation}\label{Eq:chisq_density}
f_n(x) = \frac{2^{-n/2}}{\Gamma(n/2)}x^{n/2 - 1}e^{-x/2}, \qquad x \ge 0,
\end{equation}
the density of the \emph{central} chi-square distribution with $n$ degrees of freedom. The density $f_{n,\lambda}(x)$ of the \emph{non-central} chi-squared distribution with $n$ degrees of freedom and non-centrality parameter $\lambda \ge 0$ can be expressed as an infinite weighted sum of densities of central chi-square distributions:
\[f_{n,\lambda}(x) = \sum_{i=0}^\infty{\frac{e^{-\lambda/2}(\lambda/2)^i}{i!}f_{n + 2i}(x)}, \qquad x \ge 0.\]
Finally, note that $\frac{1}{c}f_n\left(\frac{x}{c}\right)$, with $c > 0$ is the density of a gamma distribution with shape parameter $n/2$ and scale parameter $2c$. Using Lemma~\ref{Lem:conditional_chisq}, we have
\begin{align}
&\E{g_{n,T}(RV^X_n(T))} = \E{\Econd{g_{n,T}\left(RV^X_n(T)\right)}{\cL}} \notag \\
&= \E{\int_0^\infty{\sum_{i=0}^\infty g_{n,T}\left(\frac{\sigma^2}{n}x\right)\left(\frac{e^{-\lambda(T)/2}(\lambda(T)/2)^i}{i!}f_{n + 2i}(x)\right)dx}}.\label{Eq:chisq}
\end{align}
Let $\mathbb{M}$ be the counting measure that assigns mass $1$ to each integer in $\NN_0$. The last term in \eqref{Eq:chisq} can then be regarded as integrating the function
\[h_{n,T}(\omega,x,i) = g_{n,T}\left(\frac{\sigma^2}{n}x\right)\left(\frac{e^{-\lambda(T)/2}(\lambda(T)/2)^i}{i!}f_{n + 2i}(x)\right)\]
 with respect to the product measure $\PP \otimes dx \otimes \mathbb{M}$. We want to evaluate this integral as $T \to 0$. By Lemma~\ref{lem:auxrv}, $\lambda(T) \to 0$ $\PP$-almost surely. Hence we may assume that $T$ is small enough to ensure $\lambda(T) \le 1$,  almost everywhere with respect to $\PP \otimes dx \otimes M$. In this case we can estimate the integrand using the explicit form \eqref{Eq:chisq_density} of the chi-square density:
\[\left|h_{n,T}(\omega,x,i)\right| \le \norm{g_{n,T}}_\infty \frac{(x/4)^i}{i!} \left(x/2\right)^{n/2-1}e^{-x/2}.\]
Summing the right-hand side with respect to the counting measure $\mathbb{M}$ we get $\norm{g_{n,T}}_\infty  (x/2)^{n/2-1} e^{-x/4}$, which is $\PP \otimes dx$-integrable. Therefore, dominated convergence allows us to interchange limit and integration, and we obtain
\begin{align*}
\lim_{T \to 0}\E{g_{n,T}(RV^X_n(T))} &= \E{\int_0^\infty{\sum_{i=0}^\infty \lim_{T \to 0} h_{n,T}(\omega,x,i)dx}} \\
&= \int_0^\infty{g_{n,0}\left(\frac{\sigma^2}{n}x\right) f_n(x) dx}.
\end{align*}
Using the fact that a scaled chi-square distribution is a gamma distribution, the result follows.
\end{proof}

Having determined the small-time limit of option prices on both quadratic variation and realized variance, we now consider the difference between the two.

\begin{defn}\label{Def:gap}
Assume that the limiting payoff $g$ is the same for all options, i.e., there exists a function $g:\mathbb{R}_+ \to \mathbb{R}$ such that $\norm{g_T - g}_\infty \to 0$ and $\norm{g_{n,T} - g}_\infty \to 0$ as $T \to 0$ for all $n \in \NN$. Then we define
\begin{equation}\label{Eq:discretization_gap}
 \Delta_n(g) := \lim_{T \to 0} \left(\E{g_{n,T}(RV_n^X(T))} - \E{g_T(\tfrac{1}{T}[X,X]_T)}\right),
\end{equation}
and call $\Delta_n(g)$ the \emph{discretization gap}.
\end{defn}

\begin{cor}\label{Cor:discretization_gap}
Suppose the prerequisites of Theorems \ref{thm:qvlimit} and \ref{thm:rvlimit} are satisfied. Then the discretization gap $\Delta_n(g)$ from Definition~\ref{Def:gap} is given by
 \begin{equation}\label{Eq:discretization_gap_2}
\Delta_n(g) = \E{g(Y_n) - g(\sigma^2)},
 \end{equation}
 where $Y_n$ is a gamma-distributed random variable as in Theorem~\ref{thm:rvlimit}. If, in addition, the function $g$ is convex, $\Delta_n(g)$ has the following properties:
 \begin{enumerate}[(a)]
  \item $\Delta_n(g) \ge 0$ for all $n \in \NN$,\label{Item:gap_positive}
  \item $\Delta_n(g) = 0$ if and only if $\sigma^2 = 0$ or $g$ is affine-linear,\label{Item:gap_zero}
  \item $n \mapsto \Delta_n(g)$ is decreasing in $n$ and converges to $0$ as $n \to \infty$.\label{Item:gap_decreasing}
 \end{enumerate}
\end{cor}
\begin{proof}
Since $g$ is convex and $\E{Y_n} = \sigma^2$, Jensen's inequality yields $\E{g(Y_n)} \ge \E{g(\sigma^2)}$ and \eqref{Item:gap_positive} follows. Equality clearly holds if $\sigma^2 = 0$ or $g$ is affine-linear, which yields the `if'-part of \eqref{Item:gap_zero}. For the `only if'-part assume that $\sigma^2 > 0$ and that $g$ is not affine-linear. Then $g$ is \emph{strictly} convex at least on some interval $(a,b)$. Since $\sigma^2 > 0$ the interval $(a,b)$ has strictly positive measure under the law of $Y_n$ and the strict Jensen inequality implies that $\Delta_n(g) > 0$, completing the proof of \eqref{Item:gap_zero}. By \cite[Example~1.5.1e]{Stoyan1983} the gamma distributed random variables $Y_n$ are decreasing in the convex stochastic order. In particular, $n \mapsto \E{g(Y_n) - g(\sigma^2)}$ is decreasing, too. Finally $Y_n$ converges to $\sigma^2$ in distribution as $n \to \infty$ by elementary properties of the gamma distribution; hence $\lim_{n \to \infty} \Delta_n(g)=0$, showing \eqref{Item:gap_decreasing}.
\end{proof}

The above corollary has some very interesting implications.
\begin{itemize}
\item Assertion~\eqref{Item:gap_positive} shows that -- at least asymptotically for small maturity -- an option on quadratic variation is always cheaper than the option on realized variance with the same payoff, given that the payoff is convex.
\item Assertion~\eqref{Item:gap_zero} shows that the difference between the two option prices -- the discretization gap -- vanishes in two cases: The first case is when the payoff is linear; this confirms the observation of B\"uhler quoted in the introduction, and explains why for variance swaps realized variance can be substituted by quadratic variation even for short maturities. The second case in which the discretization gap vanishes is for L\'evy processes without a diffusion component, (i.e., with $\sigma^2 = 0$). This suggests that in a pure-jump L\'evy model, quadratic variation should be a good proxy for realized variance, even when pricing short-dated options with non-linear (convex) payoffs. This assertion is confirmed by our numerical examples in Section~\ref{Sec:Example}.
\end{itemize}
Assertion~\eqref{Item:gap_decreasing}, finally, is also quite intuitive. Since realized variance converges to quadratic variation as $n \to \infty$, also the discretization gap should vanish in the limit. It does, and in fact it does so monotonically in $n$.

\subsection{Applications to put and call options}\label{sec:discussion}
Let us now examine the important special cases of put and call options. More specifically, denote by $V_T$ the variance swap rate, and consider the payoffs
\begin{align*}
 &x \mapsto (kV_T - x)^+ \qquad \text{resp.} \qquad x\mapsto (x - kV_T)^+
\end{align*}
for puts resp. calls with relative strike value $k > 0$. Setting $k = 1$ yields ATM options. If the realized variance is approximated by quadratic variation, the swap rate is given by
$$V_T := \E{\tfrac{1}{T}[X,X]_T} = \sigma^2 + v^2,$$
where $v^2 = \int x^2 F(dx)$. We may apply Theorem~\ref{thm:qvlimit} to the payoff
$$g_T(x)=g_0(x)=\left(\tfrac{k}{T}\E{[X,X]_T}-x\right)^+=\left(k(\sigma^2+v^2)-x\right)^+,$$ to obtain the small-time limit for put options. For call options we use put-call parity: The difference between a call and a put option with relative strike $k$ is $(1-k)$ times the swap rate.

\begin{cor}\label{Cor:putcall_qv} Let $X$ be a square-integrable L\'evy process with L\'evy-Khintchine triplet $(b,\sigma^2,F(dx))$, and define $v^2 = \int x^2 F(dx)$. Then the following holds:
\begin{enumerate}[(a)]
\item Put options on quadratic variation satisfy
\begin{equation}\label{Eq:limit_put_qv}
 \lim_{T \to 0} \E{(k V_T - \tfrac{1}{T}[X,X]_T)^+} = \left[\sigma^2 (k-1) + v^2 k\right]^+ ,
\end{equation}
\item Call options on quadratic variation satisfy
\begin{equation}\label{Eq:limit_call_qv}
 \lim_{T \to 0} \E{\tfrac{1}{T}[X,X]_T - kV_T)^+} = v^2 + \left[\sigma^2(1 - k) - v^2 k\right]^+.
\end{equation}
\end{enumerate}
\end{cor}

Note that in the ATM case both limits coincide and are equal to $v^2$. In other words, the small-time limit of the price of an ATM option on quadratic variation is equal to the second moment of the jump measure and vanishes precisely in the Black-Scholes model. In the out-of-the-money case (puts with $k < 1$ and calls with $k > 1$) the limit may be zero even when jumps are present.\\

When the discretely sampled realized variance is used, calculations are a bit more involved. In this case, the swap rate is given by
\begin{equation}\label{Eq:swap_rate}
V^n_T := \E{RV^X_n(T)}= (\sigma^2 + v^2) +  b^2 T/n.
\end{equation}
In particular $V^n_0 := \lim_{T \to 0} V^n_T = (\sigma^2 + v^2)$. For put options Theorem~\ref{thm:rvlimit} can be applied directly, for call options we use again put-call parity.

\begin{cor}\label{Cor:putcall_rv} Let $X$ be a square-integrable L\'evy process with L\'evy-Khintchine triplet $(b,\sigma^2,F(dx))$, and define $v^2 = \int x^2 F(dx)$. Then the following holds:
\begin{enumerate}[(a)]
\item Put options on realized variance satisfy
\begin{equation}\label{Eq:limit_put}
 \lim_{T \to 0} \E{(k V^n_T - RV_n(T))^+} = \sigma^2 Q_{k,n}\left(\frac{v^2}{\sigma^2}\right) + \Big(\sigma^2 (k-1) + v^2 k\Big) R_{k,n}\left(\frac{v^2}{\sigma^2}\right),
\end{equation}
\item Call options on realized variance satisfy
\begin{multline}\label{Eq:limit_call}
 \lim_{T \to 0} \E{(RV_n(T) - kV^n_T)^+} = \\
 =v^2 + \sigma^2 Q_{k,n}\left(\frac{v^2}{\sigma^2}\right)  + \Big(\sigma^2 (k-1) + v^2 k\Big) \left\{R_{k,n}\left(\frac{v^2}{\sigma^2}\right) - 1\right\}.
\end{multline}
\end{enumerate}
The functions $Q_{k,n}(r)$ resp. $R_{k,n}(r)$ are strictly decreasing resp. increasing functions on $[0,\infty)$ given by
\[Q_{k,n}(r) = \frac{2/n}{\Gamma(n/2)}\left(\frac{n}{2}\frac{k(1 + r)}{\exp(k(1 + r))}\right)^{n/2}, \quad R_{k,n}(r) = \frac{\gamma(n/2,k(1 + r)n/2)}{\Gamma(n/2)},\]
where $\Gamma(x)$ denotes the (complete) gamma function and $\gamma(n,x)$ the lower incomplete gamma function.
\end{cor}

Note that it follows from Corollary~\ref{Cor:discretization_gap} that both \eqref{Eq:limit_put} and \eqref{Eq:limit_call} are \emph{decreasing} functions of $n$, which is illustrated by the numerical examples in Section~\ref{Sec:Example}. Also note that contrary to options on quadratic variation, the limiting value of a put or call option on realized variance is never zero, apart from the trivial case of deterministic $X$.

Having derived the small-time limit for prices of puts and calls, both on realized variance and on quadratic variation, we can now consider the difference between the two, i.e., the discretization gap introduced in Definition~\ref{Def:gap}. Simplifying the notation a bit, we write
\begin{subequations}\label{Eq:difference}
\begin{align}
 \Delta P_{k,n} &= \lim_{T \to 0} \left(\E{(k V^n_T - RV^X_n(T))^+} - \E{(k V_T - \tfrac{1}{T}[X,X]_T)^+}\right)\\
 \Delta C_{k,n} &= \lim_{T \to 0} \left(\E{(RV^X_n(T) - k V^n_T)^+} - \E{(\tfrac{1}{T}[X,X]_T - k V_T)^+}\right),
 \end{align}
\end{subequations}
which are the discretization gaps for put and call payoffs respectively. The following results can be derived from Corollary~\ref{Cor:discretization_gap} or simply by combining Corollaries \ref{Cor:putcall_qv} and \ref{Cor:putcall_rv} above.

\begin{cor}\label{Cor:diff}
Let $X$ be a square-integrable L\'evy process with L\'evy triplet $(b,\sigma^2,F(dx))$. Set $v^2 = \int x^2 F(dx)$ and define the functions $Q_{k,n}(r)$ as well as $R_{k,n}(r)$ as in Corollary~\ref{Cor:putcall_rv}. Then the following holds:
\begin{enumerate}[(a)]
 \item For $\sigma^2(k-1) + kv^2 \ge 0$, we have
  \begin{multline*}
   \Delta P_{k,n} = \Delta C_{k,n} = \sigma^2 Q_{k,n}\left(\frac{v^2}{\sigma^2}\right) + \Big(\sigma^2 (k-1) + v^2 k\Big) \left\{R_{k,n}\left(\frac{v^2}{\sigma^2}\right) - 1\right\}.
 \end{multline*}
 \item For $\sigma^2(k-1) + kv^2 \le 0$, we have
  \begin{multline*}
  \Delta P_{k,n} = \Delta C_{k,n} = \sigma^2 Q_{k,n}\left(\frac{v^2}{\sigma^2}\right) + \Big(\sigma^2 (k-1) + v^2 k\Big) R_{k,n}\left(\frac{v^2}{\sigma^2}\right).
 \end{multline*}
\end{enumerate}
\end{cor}
From Corollary~\ref{Cor:discretization_gap} it follows that $\Delta P_{n,k}$ and $\Delta C_{n,k}$ are always positive and vanish  if $\sigma^2 = 0$, i.e., in a pure-jump model. Suppose now that, in a certain L\'evy model, we can easily calculate the prices of put and call options on quadratic variation\footnote{We discuss in Section~\ref{Sec:exact} how -- and in which models -- this can be done.}. Then the expressions for $\Delta P_{n,k}$ and $\Delta C_{n,k}$ can be used as correction terms to obtain an improved approximation for the price of the corresponding option on realized variance:
\begin{app}\label{Approx:convexity_corr} Let $\Delta P_{n,k}$ and $\Delta C_{n,k}$ be given by Corollary~\ref{Cor:discretization_gap}. Then the price of a put resp. call on quadratic variation can be approximated by
\begin{subequations}\label{Eq:convexity_corr}
\begin{align}
\E{\left(kV_T^n - RV_n(T)\right)^+} &\approx \E{\left(kV_T - \tfrac{1}{T}[X,X]_T\right)^+} + \Delta P_{n,k},\\
\E{\left(RV_n(T) - kV_T^n\right)^+} &\approx \E{\left(\tfrac{1}{T}[X,X]_T- kV_T\right)^+} + \Delta C_{n,k}.
 \end{align}
\end{subequations}
\end{app}
\begin{rem}
An approximation of similar type has been proposed by Sepp \cite{sepp.10} for the Heston model.
\end{rem}

These approximations are exact in the limit $T \to 0$ (by definition of $\Delta P_{n,k}$ and $\Delta C_{n,k}$), and in the limit $n \to \infty$ (since the $\Delta$-terms vanish, and realized variance converges to quadratic variation). We can therefore expect \eqref{Eq:convexity_corr} to be good approximations for \emph{all maturities}. The numerical results in Section~\ref{Sec:Example} confirm convincingly that this is the case.

Similar approximations can of course be constructed for general payoffs $g$, using the correction term $\Delta_n(g)$ from \eqref{Eq:discretization_gap_2}. According to Corollary~\ref{Cor:discretization_gap}, the discretization gap $\Delta_n(g)$ vanishes for linear payoffs, such that it can be interpreted as a \emph{convexity correction} that corrects the basic approximation \eqref{Eq:approximation} depending on the convexity of the payoff.\\

So far, we have confined ourselves to L\'evy models. It is a natural next step to examine whether our findings remain true when passing to more general asset price models incorporating, e.g., stochastic volatility. This is done in the following section.

\section{Small-time asymptotics in semimartingale models}\label{Sec:semimartingale}

In this section, we show that -- under very mild conditions -- the small-time asymptotics of options on variance for general semimartingales coincide with those of a suitable L\'evy approximation. Thus, the results derived in the previous section can be transferred directly and no new phenomena arise. Throughout, we suppose that the log-price is given by
\begin{equation}\label{eq:log}
dX_t=b_t dt + \sigma_t dW_t + \kappa_t(x) * (N(dt,dx)-F(dx)dt), \quad X_0=0,
\end{equation}
for a standard Brownian motion $W$, a Poisson random measure $N(dt,dx)$ with absolutely continuous compensator $F(dx)dt$ (cf.\ \cite[Section II.1]{js.03} for more details), and predictable integrands $b, \sigma, \kappa$. To ensure that the log-price process $X$ is a well-defined square-integrable semimartingale, we assume that
\begin{equation}\label{eq:integrable}
\int_0^T \left(\E{b_t^2}+\E{\sigma_t^2}+\E{\int \kappa_t^2(x)F(dx)}\right) dt <\infty.
\end{equation}

\begin{rem}
The processes of the form \eqref{eq:log} comprise essentially \emph{all} semimartingales with absolutely continuous characteristics, i.e., without fixed times of discontinuity (see \cite[Theorem 14.68(a)]{jacod.79} for more details).
\end{rem}

In any reasonable application, \eqref{eq:integrable} will imply
\begin{equation}\label{eq:alwayssatisfied}
\E{\int \kappa_0^2(x)F(dx)}<\infty.
\end{equation}
In this case, the process $X$ can be approximated for small $t$ by the square-integrable L\'evy process
\begin{equation}\label{eq:approxlevy}
d\bar{X}_t=b_0 dt+\sigma_0dW_t+\kappa_0(x)*(N(dt,dx)-F(dx)dt), \quad \bar{X}_0=0,
\end{equation}
obtained from $X$ by ``freezing'' the coefficients of $X$ at time zero. Subject to weak regularity assumptions on the coefficients $b$, $\sigma$, and $\kappa$, we then have the following small-time approximation results closely related to \cite[Proposition 2.1]{muhlekarbe.nutz.10}.

\begin{lem}\label{lem:approx1}
Suppose \eqref{eq:integrable} and \eqref{eq:alwayssatisfied} hold. Then if $\E{|\sigma^2_t-\sigma^2_0|} \to 0$ and \linebreak $\E{\int |\kappa_t^2(x)-\kappa_0^2(x)|F(dx)} \to 0$ for $t\to 0$, we have
$$\lim_{T \to 0} \E{|\tfrac{1}{T}[X,X]_T-\tfrac{1}{T}[\bar{X},\bar{X}]_T|}= 0.$$
\end{lem}

\begin{proof}
By \cite[Theorems I.4.52 and II.1.8]{js.03} and Fubini's theorem, we have
\begin{align*}
&\E{|[X,X]_t-[\bar{X},\bar{X}_T|}\\
&\quad  \leq \int_0^T \E{|\sigma^2_t-\sigma^2_0|}dt+\int_0^T \E{\int|\kappa^2_t(x)-\kappa^2_0(x)|F(dx)}dt.
\end{align*}
Hence the assertion follows from the regularity assumptions on $\sigma$ and $\kappa$.
\end{proof}

 \begin{lem}\label{lem:approx2}
Suppose \eqref{eq:integrable}, \eqref{eq:alwayssatisfied} hold and assume that $\E{(b_t-b_0)^2} \to 0$ as well as $\E{(\sigma_t-\sigma_0)^2} \to 0$ and $\E{\int (\kappa_t(x)-\kappa_0(x))^2F(dx)} \to 0$ for $t \to 0$. Then, for any $n \in \mathbb{N}$, we have
$$\lim_{T \to 0}\E{\big|RV^X_n (T)-RV^{\bar{X}}_n(T)\big|}= 0,$$
for the realized variances $RV^X_n(T)$ of the log-price $X$ and $RV^{\bar{X}}_n(T)$ of the L\'evy approximation $\bar{X}$.
 \end{lem}

 \begin{proof}
 First notice that the inequalities of Cauchy-Schwarz and Minkowski imply
 \begin{align}
&\E{\Big| \sum_{j=1}^n (X_{t_j}-X_{t_{j-1}})^2-\sum_{j=1}^n (\bar{X}_{t_j}-\bar{X}_{t_{j-1}})^2\Big|} \label{eq:bound0}\\
& \quad  \leq \sum_{j=1}^n \left(\E{(X_{t_j}-\bar{X}_{t_j})^2}^{1/2}+\E{(X_{t_{j-1}}-\bar{X}_{t_{j-1}})^2)}^{1/2}\right) \notag \\
& \qquad \qquad \times \left( \E{X^2_{t_j}}^{1/2}+\E{X^2_{t_{j-1}}}^{1/2}+\E{\bar{X}^2_{t_j}}^{1/2}+\E{\bar{X}^2_{t_{j-1}}}^{1/2}\right) \notag.
\end{align}
For $t \leq T$, the Burkholder-Davis-Gundy inequality as in \cite[Theorem IV.48]{protter.05} as well as \cite[Theorems I.4.52 and II.1.8]{js.03} yield
$$\E{X_t^2} \leq C \int_0^T \left(\E{b_t^2} +\E{\sigma^2_t}+\E{\int \kappa_t^2(x) F(dx)}\right) dt,$$
for a constant $C$ which does not depend on $T$. Arguing analogously for $\bar{X}$ instead of $X$, we obtain
\begin{equation}\label{eq:bound1}
\E{X^2_{t_i}}^{1/2}+\E{X^2_{t_{i-1}}}^{1/2}+\E{\bar{X}^2_{t_i}}^{1/2}+\E{\bar{X}^2_{t_{i-1}}}^{1/2}=O(T^{1/2}) \quad \mbox{as } T \to 0.
\end{equation}
Now notice that another application of the Burkholder-Davis-Gundy inequality and \cite[Theorems I.4.52 and II.1.8]{js.03} shows that, for $t \leq T$,
\begin{align*}
&\E{(X_t-\bar{X}_t)^2}\\
&\leq C \int_0^T \left(\E{(b_t-b_0)^2}+\E{(\sigma_t-\sigma_0)^2}+\E{\int (\kappa_t(x)-\kappa_0(x))^2F(dx)}\right)dt,
\end{align*}
for a constant $C$ independent of $T$. Under the stated assumptions, it follows that
$$\E{(X_{t_i}-\bar{X}_{t_i})^2}^{1/2}+\E{(X_{t_{i-1}}-\bar{X}_{t_{i-1}})^2)}^{1/2}=o(T^{1/2}) \quad \mbox{as } T \to 0.$$
Combined with \eqref{eq:bound0} and \eqref{eq:bound1}, this proves the assertion.
\end{proof}

 For Lipschitz continuous payoffs, it now is an immediate consequence of Lemmas \ref{lem:approx1} and \ref{lem:approx2} that the small-time asymptotics for the semimartingale \eqref{eq:log} and its L\'evy approximation \eqref{eq:approxlevy} coincide. Therefore they are determined by the formulas in Theorems \ref{thm:qvlimit} and \ref{thm:rvlimit}, respectively.

 \begin{thm}\label{Thm:semimartingale_limit}
Let $X$ be a semimartingale of the form \eqref{eq:log} satisfying the prerequisites of Lemma \ref{lem:approx1} resp.\ Lemma \ref{lem:approx2}.
\begin{enumerate}[(a)]
 \item Suppose that the payoff functions $g_T: \mathbb{R}_+ \to \mathbb{R}, T \ge 0$ are continuous, uniformly bounded, and satisfy $\norm{g_T - g_0}_\infty \to 0$ as $T \to 0$. Moreover, suppose that $g_0$ is Lipschitz continuous. Then
 \[\lim_{T \to 0} \E{g_T(\tfrac{1}{T}[X,X]_T)} = g_0(\sigma^2_0).\]\label{Item:semimartin_rv}
 \item Suppose that the payoff functions $g_{n,T}: \mathbb{R}_+ \to \mathbb{R}, T \ge 0, n \in \NN$ are uniformly bounded and satisfy $\norm{g_{n,T} - g_{n,0}}_\infty \to 0 $ as $T \to 0$ for each $n \in \NN$. Moreover, suppose that the $g_{n,0}$ are Lipschitz continuous. Then
 \[\lim_{T \to 0} \E{g_{n,T}(RV_n^X(T))} = g_{n,0}(Y_n),\]
 where $Y_n$ has gamma distribution with shape parameter $n/2$ and scale parameter $2\sigma_0^2/n$.\label{Item:semimartin_qv}
\end{enumerate}
\end{thm}

\begin{proof}
We decompose
\begin{align*}
  \E{\left|g_T(\tfrac{1}{T}[X,X]_T) - g_0(\sigma_0^2)\right|} &\le \E{\left|g_T(\tfrac{1}{T}[X,X]_T) - g_0(\tfrac{1}{T}[X,X]_T)\right|}  \\ &\quad+ \E{\left|g_0(\tfrac{1}{T}[X,X]_T) - g_0(\tfrac{1}{T}[\bar{X},\bar{X}]_T)\right|} \\
   &\quad + \E{\left|g_0(\tfrac{1}{T}[\bar{X},\bar{X}]_T) - g_0(\sigma_0^2)\right|},
 \end{align*}
 where $\bar{X}$ is the approximating L\'evy process \eqref{eq:approxlevy}. The first term on the right-hand side can be bounded by $\norm{g_T - g_0}_\infty$ and thus goes to zero as $T \to 0$. The second term can be bounded by $C\,\E{\left|\tfrac{1}{T}[X,X]_T - \tfrac{1}{T}[\bar{X},\bar{X}]_T\right|}$, where $C$ is the Lipschitz constant of $g_0$. By Lemma~\ref{lem:approx1}, this term goes to $0$, too. The third term also converges to $0$, by Theorem~\ref{thm:qvlimit}, and the claim \eqref{Item:semimartin_rv} follows. Assertion \eqref{Item:semimartin_qv} is shown in the same way, substituting $\tfrac{1}{T}[X,X]_T$ by $RV^X_n(T)$ and using Lemma~\ref{lem:approx2} as well as Theorem~\ref{thm:rvlimit}.
\end{proof}

Consequently, Corollary~\ref{Cor:discretization_gap} on the discretization gap, and Corollaries~\ref{Cor:putcall_qv}, \ref{Cor:putcall_rv} and \ref{Cor:diff} on put and call options on variance also hold in the present semimartingale setting if we substitute $\sigma^2 = \sigma_0^2$ and $v^2 = \int \kappa^2_0(x) F(dx)$. In particular, we find that the discussion following Corollary~\ref{Cor:discretization_gap} can be completely transferred to the semimartingale setting, and that -- again in an asymptotic sense -- B\"uhler's statement quoted in the introduction holds generically for semimartingales with non-vanishing diffusion component, where $\sigma_0 > 0$. More specifically, for a \emph{continuous} semimartingale the small-time limit for ATM options on quadratic variation is zero, but its counterpart for realized variance is not. For semimartingales with jumps \emph{and} a continuous martingale part, both limits are non-zero, but there is a non-trivial discretization gap. In \emph{pure-jump} models however, where $\sigma=0$, the two small-time limits coincide, suggesting that quadratic variation should be a good approximation even for short maturities.

As an illustration, we show now how the prerequisites of Theorem \ref{Thm:semimartingale_limit} can be verified in some applications. For the sake of clarity, we do not strive for minimal conditions.

\begin{cor}[L\'evy driven SDEs]\label{ex:sde}
Let $f: \mathbb{R} \to \mathbb{R}$ be bounded and Lipschitz continuous and let $L$ be a L\'evy process with L\'evy exponent $\psi^L$, whose L\'evy measure $F^L(dx)$ has bounded support. Then there exists a unique strong solution $Y$ to the SDE
$$dY_t=f(Y_{t-})dL_t, \quad Y_0 \in \mathbb{R}.$$
Moreover, $S=S_0 \exp(X)$ is a martingale and the prerequisites of Theorem \ref{Thm:semimartingale_limit} are satisfied for the process
$$dX_t=dY_t-\psi^L(f(Y_{t-}))dt.$$
\end{cor}

\begin{proof}
The first part of the assertion follows from the standard existence and uniqueness theorem for SDEs as in \cite[Theorem V.6]{protter.05}, because $f$ is Lipschitz. Since the support of the L\'evy measure $F^L(dx)$ is bounded, $L$ has finite moments and exponential moments of all order by \cite[Corollary 25.8 and Theorem 25.17]{sato.99}.  Hence $S$ is a local martingale by It\^o's formula and the true martingale property is a consequence of \cite[Proposition I.4.50(c)]{js.03}. Now notice that by definition of $\psi^L$, the process $X$ is of the form \eqref{eq:log} with
\begin{gather*}
b_t=-\frac{c^L}{2}f^2(Y_{t-})-\int(e^{f(Y_{t-})x}-1-f(Y_{t-})x)F^L(dx),\\
\sigma_t=\sqrt{c^L}f(Y_{t-}), \qquad \kappa_t(x)=f(Y_{t-})x,
\end{gather*}
for the L\'evy-Khintchine triplet $(b^L,c^L,F^L(dx))$ of $L$. Denote by $M>0$ the maximum of the Lipschitz constant and the uniform bound for the function $f$. Then since $Y_{t}=Y_{t-}$ a.s., for each $t$, we have
\begin{align*}
\E{|\sigma_t^2-\sigma_0^2|} &\leq 2 M^2 c^L \E{|Y_t-Y_0|},\\
\E{\int \left|\kappa^2_t(x)-\kappa_0^2(x)\right| F^L(dx)} &\leq 2 M^2 \int x^2 F^L(dx) \E{|Y_t-Y_0|}.
\end{align*}
Likewise, since $f$ is bounded and $F^L(dx)$ has compact support, there exists a constant $C>0$ such that
$$\E{(b_t-b_0)^2} \leq 2 M^2 c^L \E{|Y_t-Y_0|} + 2 M^2 C\int x^2 F^L(dx) \E{|Y_t-Y_0|}.$$
Finally,
\begin{align*}
\E{(\sigma_t-\sigma_0)^2} &\leq 2 M^2 c^L \E{|Y_t-Y_0|},\\
\E{\int (\kappa_t-\kappa_0)^2 F^L(dx)} &\leq 2 M^2 \int x^2 F^L(dx) \E{|Y_t-Y_0|}.
\end{align*}
$Y$ is right-continuous. Combined with the Burkholder-Davis-Gundy inequality, similar arguments as above show that $Y$ is also bounded in $L^2$ on any finite interval. Hence $Y_t \to Y_0$ in $L^1$ and it follows that the conditions of Theorem \ref{Thm:semimartingale_limit} are satisfied.
\end{proof}

The next corollary of Theorem \ref{Thm:semimartingale_limit} covers many stochastic volatility models from the empirical literature as, e.g., the ones of Bates \cite{bates.96}, where $\sigma^2$ follows a square-root process, and of Barndorff-Nielsen and Shephard \cite{barndorff.shephard.01}, where $\sigma^2$ is given by a L\'evy driven Ornstein-Uhlenbeck process.

\begin{cor}[Homogeneous jumps]\label{cor:homogeneous}
Let $\kappa_t(x)=\kappa_0(x)$ be time-homogeneous, deterministic, and such that $\int \kappa_0^2(x)F(dx)$ and $\int_{\kappa_0(x)>1} e^{2\kappa_0(x)} F(dx)$ are finite. Then for
$$b_t=-\frac{\sigma_t^2}{2}-\int (e^{\kappa_0(x)}-1-\kappa_0(x))F(dx),$$
the stock price $S=S_0 \exp(X)$ is a local martingale and the conditions of Theorem \ref{Thm:semimartingale_limit} hold, if $\sigma^2$ is right-continuous and bounded in $L^2$ in some neighborhood of zero.
\end{cor}

\begin{proof}
Evidently, \eqref{eq:integrable} is satisfied for sufficiently small $T$ under the stated assumptions. Moreover, It\^o's formula shows that $S$ is a local martingale. The regularity conditions on $\kappa$ in Theorem \ref{Thm:semimartingale_limit} are trivially satisfied. The ones for $\sigma$ follow, because the processes $\sigma^2$ and $(\sigma-\sigma_0)^2$ are uniformly integrable and hence continuous in $L^1$.
\end{proof}

\section{Exact pricing methods for options on realized variance}\label{Sec:exact}

In the previous sections, we examined the small-time limits for options written on the quadratic variation and on the discretely sampled realized variance. We have also proposed a method, Approximation~\ref{Approx:convexity_corr}, to approximate the price of an option on realized variance, given that the price of the corresponding option on quadratic variation is known. We first recall in this section how to compute these prices efficiently using Fourier-Laplace methods. We then propose a new randomization approach, that allows to use similar methods to directly determine the \emph{exact} price of an option on realized variance in exponential L\'evy models, without the use Approximation~\ref{Approx:convexity_corr}. The two methods will then be compared numerically in Section~\ref{Sec:Example}. Throughout the section we assume that the log-price $X$ follows a L\'evy process with the same properties as in Section~\ref{Sec:levy}.

\subsection{Option pricing using integral transform methods}

We first recall how to price European-style options using the integral transform approach of \cite{Carr1999,raible.00}. The key assumption is the existence of an integral representation of the option's payoff function $f$ in the following sense:
$$f(x)=\int_{R-i\infty}^{R+i\infty}  p(z) e^{-zx} dz,$$
for $p: \mathbb{C} \to \mathbb{C}$ and $R>0$ such that $v \mapsto p(R+iv)$ is integrable.

\begin{example}
For a put option we have
$$ f(x)=\frac{1}{2\pi i} \int_{R-i\infty}^{R+i\infty} \frac{e^{Kz}}{z^2} e^{-zx} dz = \frac{1}{\pi} \int_0^\infty \mathrm{Re}\left(\frac{e^{K(R+iv)}}{(R+iv)^2}e^{-(R+iv)x}\right)dv,$$
for $x \geq 0$ and any $R>0$ (cf., e.g., \cite[Corollary 7.8]{carr.lee.08}).
\end{example}

In view of Fubini's theorem, the valuation of options which can be represented like this boils down to the computation of the Laplace transform of the underlying. E.g., for the put on quadratic variation we have
$$\E{(K-\tfrac{1}{T}[X,X]_T)^+}=\frac{1}{\pi T} \int_0^\infty \mathrm{Re}\left(\frac{e^{K T (R+iv)}}{(R+iv)^2}\E{\exp\left(-(R+iv)[X,X]_T\right)}\right)dv.$$
Using the put-call parity $(x-K)^+=x-K+(K-x)^+$, this leads to the analogous formula
\begin{align*}
&\E{(\tfrac{1}{T}[X,X]_T - K)^+}\\
&\quad =\E{\tfrac{1}{T}[X,X]_T}-K+\frac{1}{\pi T} \int_0^\infty \mathrm{Re}\left(\frac{e^{K T (R+iv)}}{(R+iv)^2}\E{\exp\left(-(R+iv)[X,X]_T\right)}\right)dv
\end{align*}
for calls on variance, provided that $[X,X]_T$ is integrable. Evidently, one just has to replace the normalized quadratic variation $\frac{1}{T}[X,X]$ by $RV^X_n(T)$ to come up with the corresponding formulas for options on discretely sampled realized variance. Summing up, it remains to compute the Laplace transforms of the objects of interest.

\subsection{Options on quadratic variation}
For exponential L\'evy models, the quadratic variation process $[X,X]$ also follows a L\'evy process (cf.\ \cite{carr.al.05} for the self-decomposable and \cite{Kallsen2009} for the general case). More specifically, we have the following

\begin{lem}
Suppose the $\log$-price $X$ follows a L\'evy process with L\'evy-Khintchine triplet $(b,\sigma^2,F(dx))$. Then $[X,X]$ also is a L\'evy process and its L\'evy-Khintchine triplet is given by $(\sigma^2,0,F^{[X,X]}(dx))$ relative to the truncation function $h(x)=0$, where,
$$F^{[X,X]}(G)= \int 1_G(x^2)F(dx), \quad \forall G \in \mathcal{B}.$$
\end{lem}

\begin{proof}
See \cite[Lemma 4.1]{Kallsen2009} .
\end{proof}

Combined with the L\'evy-Khintchine formula \cite[Theorem 8.1]{sato.99}, this result immediately yields the required Laplace transform.

\begin{cor}\label{Cor:psi_QV}
We have  $\E{e^{-u[X,X]_T}}=\exp[T\psi^{[X,X]}(-u)]$, for
$$\psi^{[X,X]}(-u)=\left(-cu+\int(e^{-ux^2}-1)F(dx)\right), \quad \mathrm{Re}(u) \geq 0.$$
\end{cor}

Consequently, we obtain
\begin{equation}\label{Eq:put_QV}
\E{(K-\tfrac{1}{T}[X,X]_T)^+}=\frac{1}{\pi T} \int_0^\infty \mathrm{Re}\left(\frac{e^{K T (R+iv)}}{(R+iv)^2}e^{T\psi^{[X,X]}(-(R+iv))}\right)dv
\end{equation}
for puts on quadratic variation. Likewise, for calls on quadratic variation,
\begin{multline}\label{Eq:call_QV}
\E{(\tfrac{1}{T}[X,X]_T-K)^+}\\
 =\left(\sigma^2+\int x^2 F(dx)\right) -K +\frac{1}{\pi T} \int_0^\infty \mathrm{Re}\left(\frac{e^{K T (R+iv)}}{(R+iv)^2}e^{T\psi^{[X,X]}(-(R+iv))}\right)dv,
\end{multline}
provided that $\int x^2 F(dx)<\infty$.
Some examples from the literature where the L\'evy exponent $\psi^{[X,X]}$ of $[X,X]$ can be computed in closed form are summarized in Section \ref{Sec:Example} below.

\subsection{Options on realized variance}
In this section we develop a corresponding integral transform pricing method for claims on the discrete realized variance \eqref{Eq:realized_variance} in exponential L\'evy models, without any use of approximation. The method we present will in general be of similar computational complexity as the method based on quadratic variation, especially in cases where the L\'evy exponent $\psi$ of $X$ is of more tractable form than the L\'evy exponent $\psi^{[X,X]}$ of $[X,X]$. Such cases include for example subordination-based processes like the normal inverse Gaussian or generalized hyperbolic process, see Remark \ref{rem:eff} for more details.

As in the previous section, the crucial quantity is a Laplace transform, namely
$$\E{\exp\left(-u\sum_{j=1}^{n} (X_{t_j}-X_{t_{j-1}})^2\right)}= \left( \E{\exp(-uX_{T/n}^2)}\right)^{n},$$
where the equality is due to the independence and stationarity of the increments of the L\'evy process $X$. Consequently, if the Laplace transform of the \emph{squared} process is known, the price of, e.g., puts and calls on discrete realized variance can be recovered by an inverse Laplace transform as above.

Our approach is based on the following identity: If $Z$ is a normally distributed random variable, independent of $X_t$, then using the characteristic function of the normal distribution it holds that
\begin{equation}\label{Eq:Laplace}
\E{e^{-uX_t^2}} = \E{e^{i\sqrt{2u}X_tZ}} = \E{e^{t\psi(i Z \sqrt{2 u})}},
\end{equation}
for all $u \in \mathbb{R}_+$. Note that the first expectation is taken with respect to the law of the L\'evy process $X_t$, the middle expectation with respect to the product law of $X_t$ and $Z$, and the final expectation with respect to the law of the normal random variable $Z$ only. The exchange in the order of integration is justified by the Fubini theorem, and the fact that the integrands on the left and right hand side are bounded by $1$ in absolute value. The benefit of formula \eqref{Eq:Laplace} is to replace an integration with respect to the law of the L\'evy process -- which is typically not known explicitly -- by an integration with respect to a standard normal distribution. The characteristic exponent $\psi$ which appears in the expectation on the right is in most cases analytically known and of considerably simpler form than the law of the L\'evy process.

Let us remark here that the randomization approach of formula \eqref{Eq:Laplace} can be extended to the Laplace transform of powers $|X_t|^p$ with $p \in (0,2)$, and consequently to the \emph{discrete realized $p$-variation} $\sum_j |X_{t_j} - X_{t_{j-1}}|^p$ of a L\'evy process $X$. To this end, replace the standard normal variable $Z$ by a symmetric $\alpha$-stable random variable $S_p$ with parameters $(\alpha,\beta,c,\tau) = (p,0,1,0)$ (cf. \cite[Theorem~14.15]{sato.99}). Using that $\E{e^{iwS_p}} = \exp(-|w|^p)$ we obtain
\begin{equation}\label{Eq:randomization_stable}
\E{e^{-u|X_t|^p}} = \E{e^{iu^{1/p} X_t S_p}} = \E{e^{t\psi(i S_p u^{1/p})}},
\end{equation}
for all $u \in \mathbb{R}_+$.

\subsubsection*{Remarks on Laplace Inversion}
The integral in formula \eqref{Eq:call_QV} can be considered as inverting a Laplace transform by integration along a contour in the complex plane. There are many alternatives to this inversion method, see, e.g., \cite{Davies1979} for an overview. Some of these methods only require knowledge of the Laplace transform \emph{on the positive real line}, and thus seem tailor-made for formula \eqref{Eq:Laplace} which holds -- unless further conditions are imposed -- only on $\mathbb{R}_+$. The best-known such method is probably the \emph{Post-Widder inversion formula}
\begin{equation}\label{Eq:PostWidder}
f(x) = \lim_{n \to \infty} \frac{(-1)^n}{n}\left(\frac{n + 1}{x}\right)^{n+1}\widehat{f}^{(n)}((n+1)/x), \qquad x > 0\,,
\end{equation}
where $\widehat{f}$ denotes the Laplace transform of a function $f$, and $\widehat{f}^{(n)}$ its $n$-th derivative.
The Post-Widder method suffers from slow convergence and cancellation errors, and modifications such as the Gaver-Stehfest algorithm have been introduced to improve its performance. After implementing the Gaver-Stehfest algorithm and performing some numerical tests, we observed, however, that small errors in $\widehat{f}$ -- which invariably result from the evaluation of \eqref{Eq:Laplace} -- are strongly amplified by this method and lead to huge errors in $f$, probably due to the use of very high-order derivatives in \eqref{Eq:PostWidder}. Moreover, as \cite{Davies1979} shows, inversion algorithms that evaluate the Laplace transform in the complex half-plane are in general numerically superior to algorithms that evaluate the Laplace transform only on the positive half-plane. For these reasons we decided to concentrate on the contour integration formula \eqref{Eq:call_QV}, and as a next step, to extend \eqref{Eq:Laplace} to the complex half plane $\Re(u) > 0$. Note that this extension is also necessary for other more recent Laplace inversion algorithms like the methods proposed in \cite{Abate1995}.

\subsubsection*{Extension to the complex half plane}
Extending \eqref{Eq:Laplace} to the complex half plane will not be possible without imposing some conditions on $\psi$. The following is sufficient:
\begin{cond}\label{Cond:psi}
 The characteristic exponent $\psi$ has an analytic extension from the imaginary halfline $i\mathbb{R}_+$ to the sector
\[\Lambda = \set{u \in \CC: \frac{\pi}{4} < \arg(u) < \frac{3\pi}{4} }.\]
Moreover, the extended function $\psi$ satisfies the growth bound
\begin{equation}\label{Eq:growth}
\limsup_{r \to \infty}\frac{\Re(\psi(r e^{i \theta}))}{r^2} \le 0 \quad \text{for all} \quad \theta \in \left(\tfrac{\pi}{4},\tfrac{3\pi}{4}\right)\,.
\end{equation}
\end{cond}
\begin{rem}
An analytic function satisfying the growth bound \eqref{Eq:growth} is often said to be of \emph{order $2$ and type $0$ in the sector $\Lambda$}.
\end{rem}
An elementary symmetry argument shows that given the above condition,
$\psi$ can also be analytically extended from the negative imaginary halfline $-i \mathbb{R}_+$
to the conjugate sector $\overline{\Lambda}$.
\begin{lem}\label{Lem:reflection}
 Suppose $\psi$ satisfies Condition~\ref{Cond:psi}. Then it can also be analytically extended from the halfline $-i \mathbb{R}_+$ to the conjugate sector $\overline{\Lambda}$. Overall, $\psi$ has a unique extension to the hourglass shaped region $\Lambda^{\bowtie} = \Lambda \cup \set{0} \cup \overline{\Lambda}$, which is analytic on both $\Lambda$ and $\overline{\Lambda}$ and satisfies the growth bound
\begin{equation}\label{Eq:growth2}
\limsup_{r \to \pm \infty}\frac{\Re(\psi(r e^{i \theta}))}{r^2} \le 0 \quad \text{for all} \quad \theta \in \left(\tfrac{\pi}{4},\tfrac{3\pi}{4}\right)\,.
\end{equation}
\end{lem}
\begin{proof}
Suppose that $\psi$ satisfies Condition~\ref{Cond:psi}, i.e., it is an analytic function defined on $\Lambda$. For $u \in \overline{\Lambda}$ define $\psi(u) = \overline{\psi(\overline{u})}$. On the imaginary axis, this definition agrees with the L\'evy-Khintchine representation of $\psi$. The analyticity of $\psi$ on $\overline{\Lambda}$ follows directly, e.g., by verifying the Cauchy-Riemann differential equations. The growth bound on $\Lambda^{\bowtie}$ is an immediate consequence of the construction of the extension.
\end{proof}

We can now establish the central result of this section:

\begin{thm}\label{Thm:Laplace_formula}
Let $X_t$ be a L\'evy process with characteristic exponent $\psi$ and let $Z$ be an independent standard normal random variable. Then
\begin{equation}\label{Eq:Laplace_Lsquare}
\E{e^{-uX_t^2}} = \E{e^{t\psi(iZ\sqrt{2 u})}}
\end{equation}
holds for all $u$ on the \emph{positive real line}. If $X_t$
satisfies Condition~\ref{Cond:psi}, then \eqref{Eq:Laplace_Lsquare} holds for all
$u$ in the \emph{positive half-plane} $\set{u \in \mathbb{C}: \Re(u) > 0}$, with $\psi$ denoting the unique analytic extension described in Lemma~\ref{Lem:reflection}.
\end{thm}
\begin{rem}
The square root denotes the principal branch of the complex square root function with branch cut along the negative real line.
\end{rem}

\begin{rem}\label{rem:eff}
For most L\'evy processes proposed in the literature, the L\'evy exponent $\psi$ can be computed in closed form. Hence, the evaluation of the Laplace transform of $X_T^2$ typically requires one numerical integration. The corresponding formula for the Laplace transform of $[X,X]_T$ in Corollary \ref{Cor:psi_QV} is therefore simpler, if the integral $\int (e^{-ux^2}-1)F(dx)$ can be computed in closed form. However, even if this is possible as, e.g., for CGMY processes and the models of Merton and Kou, one usually has to employ special functions (cf.\ Section \ref{Sec:Example}) such that the numerical advantage is not too big. On the other hand, e.g., for NIG or generalized hyperbolic L\'evy processes, $\int (e^{-ux^2}-1)F(dx)$ has to be evaluated using numerical quadrature, such that both formulas turn out to be of a similar complexity.
\end{rem}

\begin{proof}
Let $u \in \mathcal{H}_+ := \set{u \in \mathbb{C}: \Re(u) > 0}$. The function $u \mapsto i\sqrt{2u}$ (using the principal branch of the square root) is a single-valued analytic function on $\mathcal{H}_+$, mapping $u$ to $\sqrt{2|u|}\exp\left(\frac{i}{2}(\arg(u) + \pi)\right)$, and thus $\mathcal{H}_+$ to $\Lambda$. With the normal random variable $Z$ taking values in $\RR$ it follows that $iZ\sqrt{2u} \in \Lambda^{\bowtie}$. Let $\epsilon > 0$. Then \eqref{Eq:growth2} implies that there exists $M_\theta > 0$ such that
\begin{equation}\label{Eq:bound}
\Re(\psi(r e^{i\theta})) \le \epsilon r^2 + M_\theta, \quad \text{for} \quad r \in \RR,\quad \theta \in \left(\tfrac{\pi}{4},\tfrac{3\pi}{4}\right).
\end{equation}
Thus
\[\left| e^{\psi(iZ\sqrt{2u})} \right| = \exp \left[\mathrm{Re}\left(\psi \left(Z\sqrt{2|u|}e^{i (\arg u + \pi)/2}\right)\right)\right] \le \exp\left(\epsilon 2|u|Z^2 +M_\theta \right). \]
Note that $Z^2$ follows a chi-square distribution with one degree of freedom. The right hand side thus has a finite expectation of value $(1 - 4\epsilon |u|)^{-1/2} e^{M_\theta}$, whenever $|u| < 1/(4\epsilon)$. Since $\epsilon$ was arbitrary, it can be chosen small enough to satisfy this condition. We have shown that
\[f(u) = \E{e^{\psi(iZ\sqrt{2u})}}\]
exists for all $u \in \mathcal{H}_+$. Next we show that it is also analytic. Let $Z_n = Z \Ind{|Z| \le n}$ be a sequence of truncations of $Z$ and define
\[f_n(u) = \E{e^{\psi(iZ_n\sqrt{2u})}}.\]
Since $\psi$ is continuous on $\Lambda^{\bowtie}$, $f_n \to f$ pointwise in $\mathcal{H}_+$. Moreover, since the integrand is absolutely bounded for $u$ in compacts, each $f_n$ is analytic in $\mathcal{H}_+$ (cf. \cite[Chapter~10, Exercise~15]{Rudin1966}). Let $\mathcal K$ be a compact subset of $\mathcal{H}_+$. On $\mathcal K$ the bound \eqref{Eq:bound} can be turned into a \emph{uniform} bound
\[\Re(\psi(u)) \le \epsilon R^2 + M, \qquad u \in \mathcal K,\]
where $R$ and $M$ depend only on $\mathcal K$, and we again use the continuity of $\psi$ on $\Lambda^{\bowtie}$. By the Cauchy-Schwarz inequality, we obtain
\begin{align*}
|f(u) - f_n(u)|^2 &= \E{\exp\left(2 \Re(\psi(iZ\sqrt{2u}))\right)\Ind{|Z| > n}}^2 \\
&\le \E{\exp\left(\epsilon 4 R Z^2 + 2 M\right)}  \PP(|Z| > n) \\
&= (1 - 8 R\epsilon)^{-1/2} e^{2 M} \PP(|Z| > n),\;
\end{align*}
for all $u \in \mathcal K$. This shows that the convergence of $f_n$ to $f$ is uniform on compact subsets of $\mathcal{H}_+$. But analyticity is preserved by uniform convergence on compacts (cf. \cite[Theorem~10.27]{Rudin1966}), such that $f$ is analytic. We have now shown that both sides of \eqref{Eq:Laplace_Lsquare} are well-defined analytic functions on $\mathcal{H}_+$. Since they coincide on the positive real line, they must coincide on all of $\mathcal{H}_+$, and the proof is complete.
\end{proof}

The following example shows that Condition~\ref{Cond:psi} can not be reduced to analyticity in the sector $\Lambda$ alone:

\begin{example}
Let $X_t = N_t - t\gamma$, where $N_t$ is a Poisson process with intensity $1$, and $\gamma = e - 1$, such that $e^X$ is a martingale. The L\'evy exponent of this process is given by $\psi(u) = e^u - 1 - u\gamma$. Clearly, $\psi$ has an analytic extension to the whole complex plane and, in particular, to the sector $\Lambda$. But $\Re(\psi(re^{i\theta})) = e^{r \cos(\theta)} \cos(r\sin(\theta)) - 1 - r \gamma \cos(\theta)$, such that the growth condition \eqref{Eq:growth} is not satisfied, e.g., in the direction $\theta = 3\pi/8$. Finally the formula \eqref{Eq:Laplace_Lsquare} is \emph{not} well-defined on the whole complex half-plane $\set{u \in \mathbb{C}: \Re(u) \ge 0}$. Indeed, a tedious calculation shows that $\Re\left(\E{\left(e^{t\psi(iZ\sqrt{2 u})}\right)^+}\right)$ is infinite for, e.g.,  $t = 1$ and $u = 3/8 - i/2$, such that $\E{e^{t\psi(iZ\sqrt{2 u})}}$ does not exist.
\end{example}

Even though Theorem~\ref{Thm:Laplace_formula} fails in this simple case, Condition~\ref{Cond:psi} holds for most L\'evy processes  used in applications.

\begin{example}
Condition~\ref{Cond:psi}  is satisfied for the following L\'evy processes.
\begin{enumerate}
\item \emph{Brownian motion}: In this case, $\psi(u)= \frac{\sigma^2}{2}(u^2-u)$ is an entire function. Moreover, $ \limsup_{r \to \infty} \mathrm{Re}(\psi(re^{i\theta}))/r^2=\frac{\sigma^2}{2}\mathrm{Re}(e^{2i\theta}) \leq 0$ for all $\theta \in (\frac{\pi}{4},\frac{3\pi}{4})$.
\item The \emph{Kou} model: This jump-diffusion process corresponds to
$$\psi(u)=\mu u + \frac{1}{2} \sigma^2 u^2 + \frac{\lambda_+ u}{\nu_+ - u} -\frac{\lambda_{-} u}{\nu_{-}+u},$$
for $\lambda_{+}, \lambda_{-}, \nu_{+}, \nu_{-} \geq 0$ and $\mu \in \mathbb{R}$ determined by the martingale condition $\psi(1)=0$. Again, $\psi$ obviously admits an analytic extension to $\Lambda$ and, in addition,
$ \limsup_{r \to \infty} \mathrm{Re}(\psi(re^{i\theta}))/r^2=\frac{\sigma^2}{2}\mathrm{Re}(e^{2i\theta}) \leq 0$ for all $\theta \in (\frac{\pi}{4},\frac{3\pi}{4})$.

\item The \emph{Merton} model: For this jump-diffusion process, we have
$$\psi(u)= \mu u + \frac{\sigma^2}{2}u^2+\lambda\left[ \exp\left(\gamma u +\frac{\delta^2}{2} u^2\right)-1\right],$$
for $\sigma \geq 0$,  $\lambda, \delta > 0$, $\gamma \in \mathbb{R}$ and $\mu \in \mathbb{R}$ determined by the martingale condition $\psi(1)=0$. Consequently, $\psi$ can be analytically extended to $\Lambda$. Furthermore, since $\mathrm{Re}(\gamma r e^{i\theta}+\frac{\delta^2}{2}r^2 e^{2i\theta}) \leq 0$ for sufficiently large $r$, it follows that $ \limsup_{r \to \infty} \mathrm{Re}(\psi(re^{i\theta}))/r^2=\frac{\sigma^2}{2}\mathrm{Re}(e^{2i\theta}) \leq 0$ for all $\theta \in (\frac{\pi}{4},\frac{3\pi}{4})$.

\item \emph{NIG} processes: In this pure jump specification,
$$\psi(u)=\mu u+\delta(\sqrt{\alpha^2-\beta^2}-\sqrt{\alpha^2-(\beta+u)^2}),$$
where $\delta, \alpha > 0$, $\beta \in (-\alpha, \alpha)$ and $\mu$ is determined by the martingale condition $\psi(1)=0$. Once more, $\psi$ admits an analytic extension to $\Lambda$. Moreover,  $ \limsup_{r \to \infty} \mathrm{Re}(\psi(re^{i\theta}))/r^2= 0$ for all $\theta \in (\frac{\pi}{4},\frac{3\pi}{4})$.

\item \emph{CGMY} processes: These generalizations of the VG process correspond to
$$\psi(u)=C\Gamma(-Y)((M-u)^Y-M^Y+(G+u)^Y-G^Y)),$$
for parameters $C, G, M > 0$ and $Y < 2$. In particular, $\psi$ can be analytically extended to $\Lambda$ and $ \limsup_{r \to \infty} \mathrm{Re}(\psi(re^{i\theta}))/r^2= 0$ for all $\theta \in (\frac{\pi}{4},\frac{3\pi}{4})$.
\end{enumerate}
\end{example}

Based on these examples and the above counterexample we conjecture that Condition~\ref{Cond:psi} is related to the absolute continuity or smoothness of the L\'evy measure.

\section{Numerical Illustration}\label{Sec:Example}
\noindent
We now consider three numerical examples. First, we take a look at the Black-Scholes model, then we turn to the pure-jump CGMY model. Finally, we also consider the jump-diffusion model of Kou.

\begin{figure}[htbp]
\begin{center}
\includegraphics[width=11cm,height=6cm]{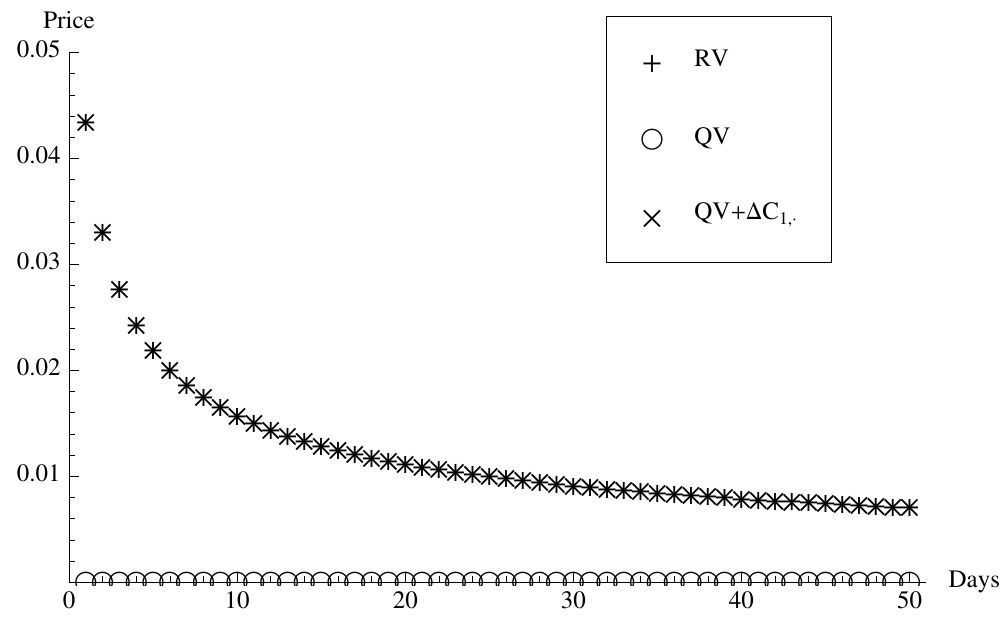}
\caption{ATM call prices on realized variance, ATM call prices on quadratic variation resp.\ convexity corrected ATM call prices on quadratic variation in the Black-Scholes model.}\label{Fig:1}
\end{center}
\end{figure}

In the Black-Scholes model the distribution of realized variance is known explicitly; it is the non-central chi-square distribution. Thus neither Approximation~\ref{Approx:convexity_corr} nor our exact method from Section~\ref{Sec:exact} would be needed to compute prices of options on realized variance. Nevertheless, the Black-Scholes model can serve as a first test case to compare our two methods. Note that Approximation~\ref{Approx:convexity_corr} amounts in this case to approximating the (true) noncentral chi-square distribution of realized variance by a \emph{central} chi-square distribution. For a volatility parameter of $\sigma=0.3$, ATM call prices on realized variance, ATM call prices on quadratic variation and ATM call prices on quadratic variation corrected by the discretization gap $\Delta C_{1,\cdot}$ are depicted in Figure \ref{Fig:1} for maturities up to $50$ days.

Evidently, the prices of ATM calls on realized variance cannot be distinguished from the convexity corrected approximation \eqref{Eq:convexity_corr} by eye, showing that Approximation~\ref{Approx:convexity_corr} performs remarkably well.

 Also note that the prices of calls on the realized variance and of convexity corrected calls on quadratic variation converge to the prices of calls on quadratic variation (which are zero) for increasing maturity, but the rate appears to be even slower than in the results for the Heston model reported in \cite{Buhler2006}.  In particular, using quadratic variation as a proxy for realized variance does not work well here, unless one uses the convexity correction \eqref{Eq:convexity_corr}.

\begin{figure}[htbp]
\begin{center}
\includegraphics[width=11cm,height=6cm]{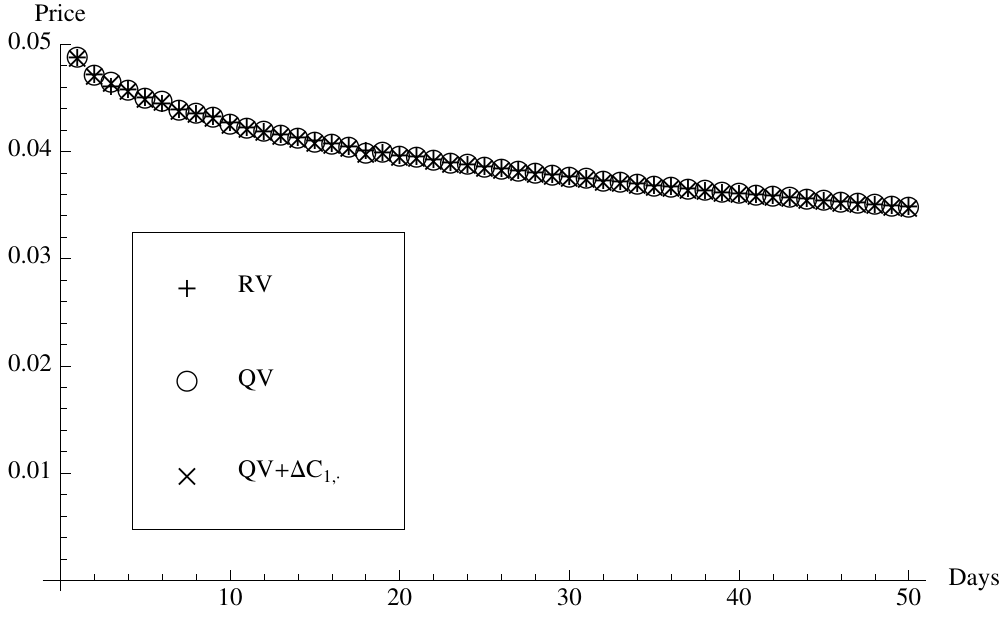}
\caption{ATM call prices on realized variance, ATM call prices on quadratic variation resp.\ convexity corrected ATM call prices on quadratic variation in the CGMY model.}\label{Fig:2}
\end{center}
\end{figure}

Next, we turn to the pure-jump CGMY process. By \cite[Section 4]{carr.al.05},
\begin{align*}
&\psi^{[X,X]}(u)=C\Bigg(\left(\frac{2u}{Y}-\frac{M^2}{Y(1-Y)}\right)I(2-Y,M,-u)\\
&\quad +\left(\frac{2u}{Y}-\frac{G^2}{Y(1-Y)}\right)I(2-Y,G,-u)+\frac{2uM}{Y(1-Y)}I(3-Y,M,-u)\\
&\quad +\frac{2uG}{Y(1-Y)}I(3-Y,G,-u)+\frac{M^Y+G^Y}{Y(1-Y)}\Gamma(2-Y)\Bigg),
\end{align*}
where
\begin{equation}\label{Eq:Ifunction}
I(\kappa,\nu,\tau):=2^{-\kappa}\tau^{-\kappa/2}\Gamma(\kappa)U\left(\frac{\kappa}{2},\frac{1}{2},\frac{\nu^2}{4\tau}\right)
\end{equation}

for the \emph{confluent hypergeometric $U$-function} $U$. We use the calibrated (yearly) parameters
$$C=0.3251, \quad  G=3.7103,\quad M=18.4460,\quad Y=0.6029,$$
from \cite[Table 1]{carr.al.05}.  Corollary \ref{Cor:psi_QV}, Approximation \ref{Approx:convexity_corr}, and Theorem \ref{Thm:Laplace_formula} then lead to the results in Figure \ref{Fig:2}.

Evidently, the three price curves cannot be distinguished by eye. In particular, quadratic variation seems to serve as an excellent proxy for realized variance at all maturities here, which drastically differs from the results reported for the Heston model in \cite{Buhler2006}, and also from the results for the Black-Scholes model shown in Figure~\ref{Fig:1}. This reflects the fact that the discretization gap vanishes for pure-jump models according to Corollary~\ref{Cor:diff}. In particular, the convexity correction term $\Delta C$ is zero in this case.

\begin{figure}[htbp]
\begin{center}
\includegraphics[width=11cm,height=6cm]{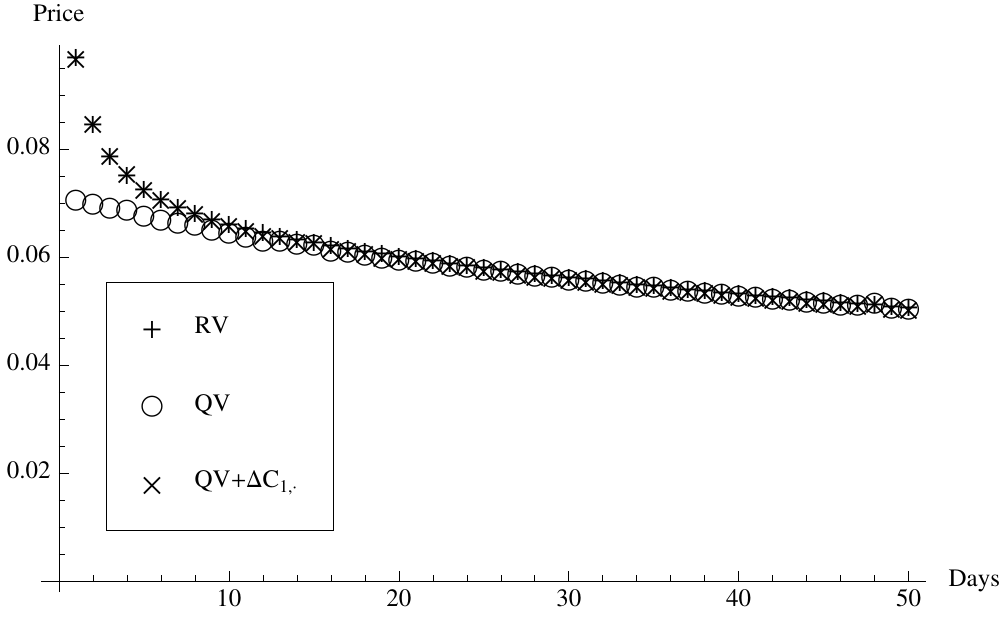}
\caption{ATM call prices on realized variance, ATM call prices on quadratic variation resp.\ convexity corrected ATM call prices on quadratic variation in the in the Kou model for $\sigma=0.3$.}\label{Fig:3}
\end{center}
\end{figure}

As a third example we consider the model of Kou, which includes both jumps and a Brownian component. Similarly as above, the L\'evy exponent $\psi^{[X,X]}$ of $[X,X]$ can again be expressed in terms of the confluent hypergeometric $U$-function:
$$\psi^{[X,X]}(u)=\sigma^2 u + \lambda_{+} (\nu_+ I(1,\nu_+,-u)-1)-\lambda_{-}(\nu_{-}I(1,\nu_{-},-u)-1),$$
with the function $I$ from \eqref{Eq:Ifunction}.  In this jump-diffusion model, we have a non-vanishing discretization gap by Corollary \ref{Cor:diff}. Using the calibrated yearly parameters
$$ \sigma=0.3, \quad \lambda_+=0.5955, \quad \nu_+=16.6667, \quad \lambda_{-}=3.3745, \quad \nu_{-}=10,$$
 from \cite[Section 7.3]{sepp.03}, we obtain the numerical results depicted in Figure \ref{Fig:3}.

\begin{figure}[htbp]
\begin{center}
\includegraphics[width=11cm,height=6cm]{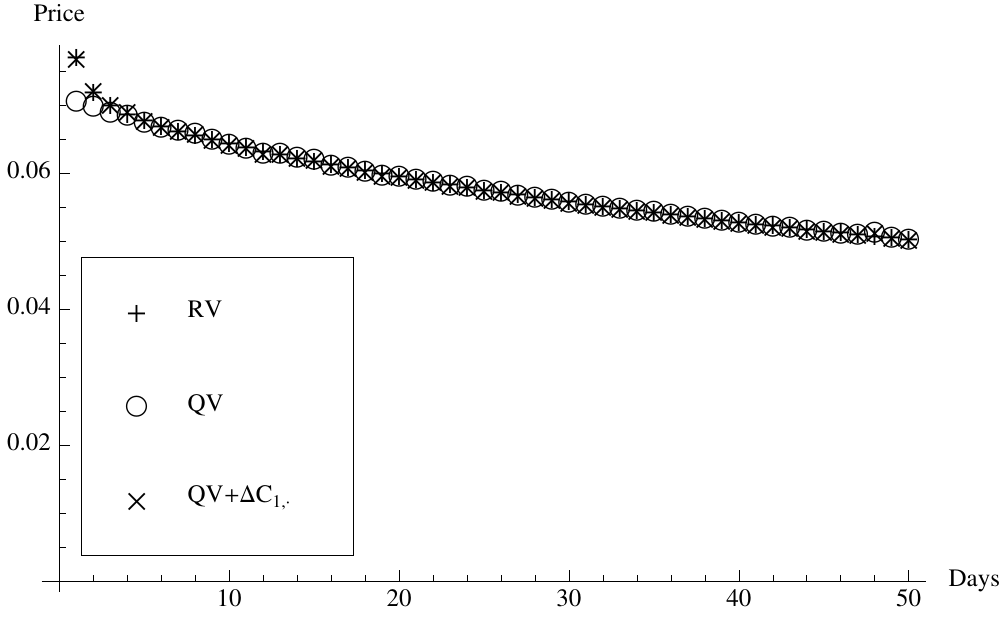}
\caption{ATM call prices on realized variance, ATM call prices on quadratic variation resp.\ convexity corrected ATM call prices on quadratic variation in the in the Kou model for  $\sigma=0.2$. }\label{Fig:kou2}
\end{center}
\end{figure}

Again, first notice that the convexity corrected call prices on quadratic variation almost perfectly match the exact call prices on realized variance. Next, note that whereas quadratic variation appears to serve as a much better approximation than in the Black-Scholes model here, the discretization gap is still significant for short maturities, unlike in the pure-jump CGMY model. By Corollary~\ref{Cor:diff} we would expect the gap to shrink for smaller values of $\sigma$. The effect of reducing $\sigma$ to $0.2$, while keeping all other parameters the same is shown in Figure \ref{Fig:kou2} and agrees with this prediction.

\section{Conclusions and Outlook}
We have proposed two different methods to calculate prices of options on realized variance, that improve upon the standard approximation by quadratic variation.

The first method, Approximation~\ref{Approx:convexity_corr}, was found to work very well for ATM options in three different L\'evy models with and without jumps. By the results in Section 3, it is also possible to apply the same approach to more general models, provided that prices of options on quadratic variation can still be computed efficiently. Hence one objective for future research will be to test its numerical performance for stochastic volatility models with and without jumps. For affine stochastic volatility models (see for example \cite{Kallsen2009,K2008a}), a class which includes the Heston model, the SVJ and SVJJ models of \cite{Gatheral2006} and most time-change based stochastic volatility models, the results of \cite{Kallsen2009} could be used as a starting point. Since our approximation method ``freezes'' the stochastic volatility at time zero, one would expect it to perform worse for stochastic volatility models. On the other hand, Sepp \cite{sepp.10} has obtained encouraging results for the Heston model with a similar method.

Our second method, the exact Fourier-Laplace approach from Section~\ref{Sec:exact}, could also be possibly extended to stochastic volatility models. Again the class of affine stochastic volatility models seems particularly suitable, since in such models the log-price $X$ and the stochastic variance process $V$ have a joint conditional characteristic function of the form
\begin{equation}\label{Eq:affine}
\Econd{e^{uX_t + wV_t}}{\cF_h} = \exp\Big(\phi(t-h,u,w) + V_h \psi(t-h,u,w) + X_h u\Big),
\end{equation}
for $u,w \in i\RR$. It seems thus possible to use a \emph{conditional} version of the identity \eqref{Eq:Laplace_Lsquare} in each time-step between business days, and to use the special form of \eqref{Eq:affine} to convert this conditional identity into a recursive algorithm for the computation of the Laplace transform of realized variance. The delicate point is to find analyticity conditions analogous to Condition~\ref{Cond:psi} that allow to extend the identity to the positive half-plane $\set{u \in \mathbb{C}: \Re(u) \ge 0}$. A rigorous analysis of the necessary technical conditions as well as an efficient numerical implementation for this case is also deferred to future research.

\appendix
\section{Proof of Lemma~\ref{lem:auxqv} and \ref{lem:auxrv}}
\begin{proof}
We first show Lemma~\ref{lem:auxqv}. Lemma~\ref{lem:auxrv} follows then after minor modifications of the proof.
Let $(T_m)_{m \in \mathbb{N}}$ be a sequence converging to zero and let $\epsilon >0$. For sufficiently small $T_m$, we have
\begin{equation}\label{eq:est1}
\E{\frac{[L,L]_{T_m}}{T_m} \wedge 1} \leq \E{\frac{[L,L]_{T_m }\wedge \epsilon}{T_m}}.
\end{equation}
Now notice that $[L,L]$ is a L\'evy process with triplet $(0,0,\int 1_{\cdot}(x^2)F(dx))$ relative to the truncation function $h(x)=0$, which can be used because $[L,L]$ is of finite variation. Hence it follows from \cite[Formula (5.8)]{Jacod2007} that
$$\lim_{T_m \to 0} \E{\frac{[L,L]_{T_m} \wedge \epsilon}{T_m}} = \int (x^2 \wedge \epsilon) F(dx).$$
Together with \eqref{eq:est1}, this implies $\E{\frac{[L,L]_{T_m}}{T_m} \wedge 1} \to 0$, because $\epsilon$ was arbitrary. As $T_m \downarrow 0$,  $[L,L]_{T_m}/T_m$ thus converges to zero in probability, and hence a.s.\ along a further subsequence $T_{m'_k}$. This shows that on a set of probability one, the sequence $[L,L]_{T_m}/T_m$ has the cluster point $0$ and, moreover, that $0$ is the unique cluster point. Therefore $[L,L]_{T_m}/T_m \to 0$ a.s. and by the right-continuity of $[L,L]_T$ the claim follows.\\
To show Lemma~\ref{lem:auxrv}, substitute $[L,L]_{T_m}$ by $L_{T_m}^2$. Now note that $L$ is a L\'evy process with triplet $(0,0,F(dx))$ relative to the truncation function $h(x)=x$, which can be used because $L$ is integrable.  Again using \cite[Formula (5.8)]{Jacod2007}, the above arguments show $\tfrac{1}{T}L_T^2 \to 0$ a.s. By independence and stationarity of the increments of $L$, it follows that $RV_n^L(T) \to 0$ a.s.
\end{proof}

\end{document}